\documentclass{edp-jp4}
\usepackage{graphicx}

\begin{document}

\title{Phase ordering induced by defects in chaotic bistable media.}

\author{C. Echeverria}\address{Laboratorio de F\'isica Aplicada Computacional, Universidad Nacional Experimental del T\'achira, San Crist\'obal, Venezuela.}
\author{K. Tucci}\address{SUMA-CeSiMo, Universidad de Los Andes, M\'erida, M\'erida 5251, Venezuela.}
\author{M. G. Cosenza}\address{Centro de F\'isica Fundamental, Universidad de Los Andes, M\'erida, M\'erida 5251, Venezuela.}

\maketitle

\begin{abstract}
The phase ordering dynamics of coupled chaotic bistable maps on lattices with defects
is investigated. The statistical properties of the system are characterized
by means of the average normalized size of spatial domains of equivalent spin variables
that define the phases. It is found that spatial defects can induce the formation of domains
in bistable spatiotemporal systems. The minimum distance
between defects acts as parameter for a transition
from a homogeneous state to a heterogeneous regime where two phases coexist  
The critical exponent of this transition also exhibits
a transition when the coupling is increased,
indicating the presence of a new class of domain where both phases
coexist forming a chessboard pattern.
 \end{abstract}
Coupled map lattices constitute fruitful and computationally efficient models for
the study of a variety of dynamical processes in spatially distributed systems
\cite{Kaneko_Mag}. The discrete-space character of coupled map systems makes them
specially appropriate for the investigation of spatiotemporal dynamics on nonuniform
and on complex networks \cite{CK1,GCR1,CT1,JA1}.

There has been recent interest in the study of the phase-ordering properties
of systems of coupled chaotic maps and their relationship with Ising models in 
statistical physics \cite{TCA1,MH1,LC1,KLC1,WLH1,APS1,SJK1}. These works have 
generally assumed the phase competition dynamics taking place on a uniform
space; however, in many physical situations the medium that supports
the dynamics can be nonuniform due to the intrinsic heterogeneous nature of
the substratum such as porous or fractured media, or it may arise from random 
fluctuations in the medium. This paper investigates the process of phase ordering in coupled 
chaotic maps on a lattice with defects as a model for studying this phenomenon on nonuniform media.

We consider a system of coupled maps defined on a two-dimensional square lattice of size $N = L \times L$
with periodic boundary conditions and having randomly distributed defects, as shown in Fig.~1. 
A defect is a non-active site, i.e., a site that possesses no dynamics. The density of defects in 
the lattice is characterized in terms of the minimum Euclidean distance $d$ between defects. 
The dynamics of the diffusively coupled map system is described by
\begin{equation}
\label{CML}
x_i(t+1) = (1-\epsilon)f(x_i(t)) + \frac{\epsilon}{{\cal{N}}_i} \sum_{j\in {\nu_i}} f(x_j(t))\;,
\end{equation}
where $x_i(t)$ is the state of an active site $i$ ($i=1,\ldots,N$) at time $t$, $\nu_i$
is the set of the nearest neighbors of site $i$ and ${\cal{N}}_i \in \{1,2,3,4\}$ is
the cardinality of this set, $\epsilon$ measures the coupling strength, and $f(x(t))$ is a chaotic map that expresses 
the local bistable dynamics \cite{MH1,LC1},
\begin{equation}
f(x) = \left\lbrace
\begin{array}{ccl}
-2\mu/3-\mu x, & \mbox{if} & x \in [-1,-1/3] \\
\mu x, & \mbox{if} & x \in [-1/3,1/3] \\
2\mu/3 - \mu x, & \mbox{if} & x \in [1/3,1] \;.
\end{array}\right. 
\end{equation} 
For $\mu \in (1,2)$ the map has two symmetric chaotic band attractors, one with values 
$x_i(t) >0$ and the other with $x_i(t)<0$, separated
by a gap about the origin. Then the local states have two well defined 
symmetric phases that can be characterized by spin variables
defined as the sign of the state at time $t$, $\sigma_i(t) = \mbox{sign}(x_i(t))$. 

\begin{figure}[htb]
\centerline{\rotatebox{90}{\resizebox{0.18\textwidth}{!}{\includegraphics{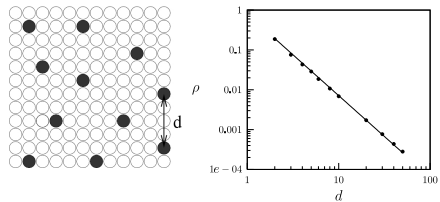}}}}
\caption{(Left) Spatial support of the system, showing active sites ($\circ$) and defects ($\bullet$). 
Defects are randomly placed in such a way that their density distribution is maximum for a given minimum 
distance $d$ between defects. 
(Right) The defect density scales as $\rho=0.625 d^{-2}$.
\label{fig:Subst_Rrhod}}
\end{figure}

We fix the local parameter at $\mu=1.9$ and set the initial conditions as follows: 
one half of the active sites are randomly
chosen and assigned random values uniformly distributed
on the positive attractor while the other half are similarly 
assigned values on the negative attractor. If the number of active sites
is odd, then the state of the remaining site is
assigned at random on either attractor. 

\begin{figure}[htb]
\centerline{\rotatebox{270}{\resizebox{0.3\textwidth}{!}{\includegraphics{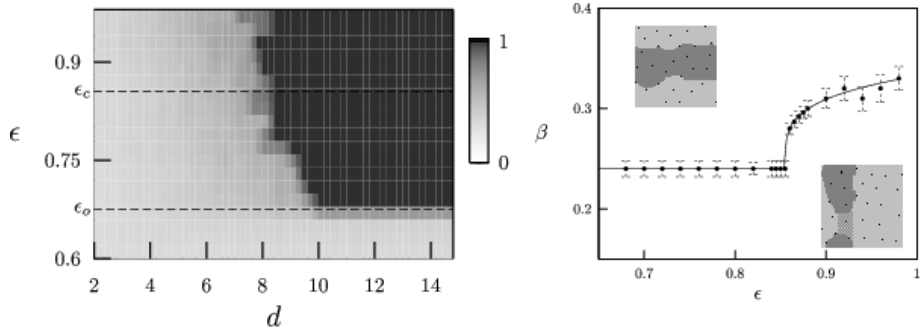}}}}
\caption{(Left) Normalized domain size $R$ as a function of $\epsilon$ and $d$. 
Dark color represents homogeneous (single-phase) states and light color corresponds to
heterogeneous (two-phase) states. (Right) Critical exponent $\beta$ vs. coupling $\epsilon$.
The two inserts show typical frozen configurations of the system for
$\epsilon < \epsilon_c$ (left) and for $\epsilon > \epsilon_c$ (right).
Dark and light colors represent each phase and black dots correspond to defects.
\label{fig:RvsEpsilon_d}}
\end{figure}

In regular lattices ($\rho=0$) phase growth occurs for values  $\epsilon > \epsilon_o$, where $\epsilon_o = 0.67$ \cite{LC1}. In contrast, in the defective medium there exist a minimum value of $\rho$ for which the domains formed by the two phases reach a frozen configuration for all values of the coupling $\epsilon$.
To characterize the phase ordering properties of the system Eq. (\ref{CML}) we use the
normalized size of the phase domains, averaged over $50$ realizations,
as an order parameter given by
\begin{equation}
R = \lim_{t \rightarrow \infty}\frac{1}{N} \sum_{r=1}^{L/2}\sum_{i,j} 
\delta_{r_{ij},r} \delta_{\sigma_i(t),\sigma_j(t)} \;,
\end{equation}
where $r_{ij}$ is the Euclidean distance between nodes $i$ and $j$.
Figure ~\ref{fig:RvsEpsilon_d} (left) shows the variation of $R$ in the space of parameters $(\epsilon ,d)$. Note that the system is heterogeneous ($R \rightarrow 0$), with coexistence of
the two phases
for values $\epsilon < \epsilon_0$. When $\epsilon > \epsilon_0$ 
two regions appear in the phase diagram: an heterogeneous regime for small values of $d$ (large $\rho$) and 
an single-phase, homogeneous state ($R=1$), for large values of $d$ (small $\rho$). 

The transition between these two types of behaviors induced by the presence of defects
for a given value of
$\epsilon > \epsilon_0$ can be described by the scaling relation
$R \sim (d_c - d)^{\beta(\epsilon)}$, where $d_c(\epsilon)$ is the threshold value of
the minimum distance at which 
the transition occurs, and $\beta$ is a critical exponent.

Figure 2 (right) shows how the behavior of the critical exponent $\beta$ varies with $\epsilon$.
There is a critical value of the coupling $\epsilon_c \approx 0.855$ at which
$\beta$ changes from being a constant value of $\beta \approx 0.24$ to follow a scaling
relation $\beta \sim (\epsilon-\epsilon_c)^{\gamma}$, where
$\gamma \approx 0.25$ is a critical exponent characterizing the transition.
This transition is related to the emergence of a new type of configuration of domains 
induced by defects in the system, where coexistence of alternating sites of the two phases takes place forming a chessboard pattern. It should be noticed that continuous variation of the values of critical exponents may also occur in some statistical mechanics models showing phase separation dynamics~\cite{ACL1,PV1}.

In summary, we have found that spatial defects can induce the formation of domains
in bistable spatiotemporal systems. The minimum distance
between defects acts as a parameter for the transition
from a homogeneous to a heterogeneous regime. 
The critical exponent of this transition $\beta$ exhibits
a second order transition when the coupling is increased,
indicating the presence of a new class of domain where both phases
alternatively coexist.

This work was supported in part by grants
04-006-2004 from Decanato de Investigaci\'on of  UNET, F-2002000426 from FONACIT, and 
I-886-05-02-A from CDCHT of Universidad de Los Andes, Venezuela.

\end{document}